\documentclass[12pt]{article}
\usepackage{%
pstricks,%
pst-plot,%
epsf%
}           
\usepackage{cite}
\usepackage{array,dcolumn}  
\usepackage{axodraw}
\newcommand{\hs}{\hspace*{0.5cm}}
\newcommand{\be}{\begin{equation}}
\newcommand{\ee}{\end{equation}}
\newcommand{\bea}{\begin{eqnarray}}
\newcommand{\eea}{\end{eqnarray}}
\newcommand{\baa}{\begin{eqnarray*}}
\newcommand{\eaa}{\end{eqnarray*}}
\newcommand{\bary}{\begin{array}}
\newcommand{\eary}{\end{array}}

\begin{document}
\begin{flushright}
APCTP-2000-004\\
\end{flushright}
\vspace*{1cm}
\begin{center}
{\large \bf   RARE KAON  DECAY $K^+ 
\rightarrow \pi^+ \nu \bar{\nu}$\\
IN  $\mbox{SU(3)}_C \otimes \mbox{SU(3)}_L
\otimes \mbox{U(1)}_N$ MODELS}\\
\vspace*{1cm}

{\bf $\mbox{Hoang  Ngoc Long}^{a,b}$,
$\mbox{Le Phuoc Trung,}^{c}$} and 
{\bf $\mbox{Vo Thanh Van}^{b}$}\\

\vspace*{0.5cm}

{\it $^a$ Asia Pacific Center for Theoretical Physics, 
Seoul 135 - 080, Korea}\\
{\it $^b$ Institute of 
Physics, NCST, P. O. Box 429, Bo Ho, Hanoi 10000, Vietnam }\\
{\it $^c$ HCM city Branch, Institute of 
Physics, Ho Chi Minh city, Vietnam }

\vspace*{1cm}

{\bf Abstract}\\

\end{center}
\hs The rare kaon decay  $K^+ \rightarrow 
\pi^+ \nu \bar{\nu}$
is considered in the framework of the  models based on 
the $\mbox{SU(3)}_C \otimes \mbox{SU(3)}_L \otimes
\mbox{U(1)}_N$ (3 - 3 - 1) gauge group.
It is shown that a lower bound of the  $Z'$ 
mass in the 3 - 3 - 1 
model with right-handed neutrinos at a value of 
3 TeV is derived, 
while that in the minimal version -- 1.7 TeV.\\

PACS number(s):12.10.-g, 12.60.-i, 13.20.Eb, 
13.25.Es, 14.60.St.\\

Keywords: Extended gauge models, rare kaon decays, 
extra neutral gauge boson $Z'$.
\vspace{1cm}

\noindent

\setcounter{footnote}{0}
\vspace*{0.5cm}

{\large\bf I. Introduction}\\[0.3cm]
\hs The kaon is the lightest hadron having 
a non-zero strangeness
quantum number. Due to the weak interactions
the kaon decays weakly into states with zero strangeness,
containing pions, photons and/or leptons.  
The physics of kaons has played a major role in the development
of particle physics. The concept of strangeness, with its
implications for the quark model, the discovery of $P$
and $CP$ violation and the
GIM mechanism have all emerged from the study of K mesons.
Today, rare kaon decays continue to be an active field
of investigation~\cite{review}.
Flavour changing neutral currents (FCNC) are completely
suppressed at the tree level by the GIM mechanism 
in the standard model (SM).
In second or higher order interactions, this suppression 
is not complete due to the different quark masses~\cite{il}.

\hs The first experimental evidence for atmospheric neutrino
oscillations and consequently a non-zero neutrino mass observed
at the SuperKamiokande Collaboration calls for the SM
extension. Among the possible models,
those based on the $\mbox{SU(3)}_C \otimes \mbox{SU(3)}_L
\otimes \mbox{U(1)}_N$ (3 -- 3 -- 1) gauge
group~\cite{pp,fr,flt,hnl}, contain
a number of intriguing features:
First, the models predict three families of quarks and
leptons if the anomaly free condition on $ \mbox{SU(3)}_L
\otimes \mbox{U(1)}_N$ and the QCD asymptotic freedom are
imposed. Second, the Peccei-Quinn
symmetry naturally occurs in these models~\cite{pal}. 
The third interesting point is that
one generation of quarks is treated differently
from two others. This could lead to a natural 
explanation for the unbalancingly heavy top
quark.  This family nonuniversality leads also
to the FCNC  by the $Z'$ currents
at the tree level~\cite{dpp,lv}. 
Finally, the 3 - 3 - 1 models predict new physics
at a scale only slightly above the SM scale
(a few TeVs)~\cite{dpp,lv,dng,jj}.

\hs In this work we consider the implications of  main two
3 - 3 - 1 models for the rare  $K^+ \rightarrow \pi^+  \nu   
\bar{\nu}$ decay, and our aim  is to get a 
bound on the $Z'$ mass.
\\[0.3cm]

{\large\bf II.  The rare kaon decay   $K^+ \rightarrow 
\pi^+  \nu   \bar{\nu}$ in 3 - 3 - 1 models}\\[0.3cm]
{\it A.   The decay in 3 -- 3 -- 1 model with 
right-handed neutrinos}\\[0.1cm]
\hs We first recapitulate the basic elements of the model.
The leptons in this model are arrangeed into  triplets, 
with the third member being 
a right-handed neutrino~\cite{flt,hnl}:
\be
f^{a}_L = \left( \begin{array}{c}
               \nu^a_L\\ e^a_L\\ (\nu^c_L)^a
\end{array}  \right) \sim (1, 3, -1/3), e^a_R\sim (1,
1, -1),
\label{l2}
\ee
where $ a = 1, 2, 3$ is the family index.

\hs The first two families of quarks are in antitriplets 
and the third one is in a triplet:
\be
Q_{iL} = \left( \begin{array}{c}
                d_{iL}\\-u_{iL}\\ D_{iL}\nonumber\\
 \end{array}  \right) \sim (3, \bar{3}, 0),\nonumber
\ee
\[ u_{iR}\sim (3, 1, 2/3), d_{iR}\sim (3, 1, -1/3),
D_{iR}\sim (3, 1, -1/3),\ i=1,2,\]
\be
 Q_{3L} = \left( \begin{array}{c}
                 u_{3L}\\ d_{3L}\\ T_{L}
                \end{array}  \right) \sim (3, 3, 1/3),\nonumber
\label{q}
\ee
\[ u_{3R}\sim (3, 1, 2/3), d_{3R}\sim (3, 1, -1/3), T_{R}
\sim (3, 1, 2/3).\]
The  gauge bosons in this model are the photon (A), $Z, Z', W^\pm,
Y^\pm$ and complex neutral bosons $X^0$ and $X^{*0}$,
\bea
\sqrt{2}\ W^+_\mu &=& W^1_\mu - iW^2_\mu ,
\sqrt{2}\ Y^-_\mu = W^6_\mu - iW^7_\mu ,\nonumber\\
\sqrt{2}\ X_\mu^o &=& W^4_\mu - iW^5_\mu,\nonumber\\
A_\mu  &=& s_W  W_{\mu}^3 + c_W\left(-
\frac{t_W}{\sqrt{3}}\ W^8_{\mu}
+\sqrt{1-\frac{t^2_W}{3}}\  B_{\mu}\right),
\nonumber\\
Z_\mu  &=&  c_W  W^3_{\mu} - s_W\left(
-\frac{t_W}{\sqrt{3}}\ W^8_{\mu}+
\sqrt{1-\frac{t_W^2}{3}}\  B_{\mu}\right),  \\
Z'_\mu &=& \sqrt{1-\frac{t_W^2}{3}}\
W^8_{\mu}+\frac{t_W}{\sqrt{3}}\ B_{\mu},\nonumber
\label{apstat1}
\eea
where we use the following notations:
$s_W \equiv \sin \theta_W $ and $t_W \equiv 
\tan \theta_W$.
The {\it physical} states are a mixture of $Z$ and $Z'$:
\bea
Z_1  &=&Z\cos\phi - Z'\sin\phi,\nonumber\\
Z_2  &=&Z\sin\phi + Z'\cos\phi,\nonumber
\eea
where $\phi$ is a mixing angle.

\hs The interactions among fermions and  $Z_1, Z_2$  
are given by
\begin{eqnarray}
{\cal L}^{NC}&=&\frac{g}{2c_W}\left\{\bar{f}
\gamma^{\mu} 
[a_{1L}(f)(1-\gamma_5) + a_{1R}(f)(1+\gamma_5)]f 
Z^1_{\mu}\right.\nonumber\\
             & &+ \left.\bar{f}\gamma^{\mu} 
[a_{2L}(f)(1-\gamma_5) + a_{2R}(f)(1+\gamma_5)]f 
Z^2_{\mu}\right\},
\label{nc}
\end{eqnarray}
where
\begin{eqnarray}
a_{1L,R}(f) &=&\cos\phi\ [T^3(f_{L,R})-s_W^2 Q(f)]
\nonumber\\
 & &- c_W^2\left[\frac{3N(f_{L,R})}{(3-4s_W^2)^{1/2}}
-\frac{(3-4s_W^2)^{1/2}}{2c^2_W}Y(f_{L,R})\right]
\sin\phi,\nonumber\\
a_{2L,R}(f)&=& c_W^2\left[\frac{3N
(f_{L,R})}{(3-4s_W^2)^{1/2}}
-\frac{(3-4s_W^2)^{1/2}}{2c^2_W}Y(f_{L,R})\right]
\cos\phi\nonumber\\
           & &+ \sin\phi\ [T^3(f_{L,R})-s_W^2 Q(f)],
\label{vaz}
\end{eqnarray}
where $T^3(f)$ and $Q(f)$ are, respectively, the third 
component of the weak isospin and the 
charge of the fermion $f$. 
The mixing angle $\phi$ is constrained 
to be very small~\cite{hnl}
$-2.8 \times 10^{-3} \leq \phi \leq 1.8 \times 10^{-4}$
and can therefore  be neglected.

\hs Because one family of 
left-handed quarks is treated 
differently from the  other two, the $N$ charges for 
left-handed quarks are also different (see Eq.~(\ref{q})). 
Therefore, the flavour-changing neutral current $ Z'$ 
occurs through a mismatch between weak and mass 
eigenstates.
We diagolize mass matrices by three biunitary 
transformations  
\begin{eqnarray}
U'_L & = & V_L^U U_L,\   U'_R = V_R^U U_R,\nonumber\\
D'_L & = & V_L^D D_L,\   D'_R = V_R^D D_R,
\label{tran}
\end{eqnarray}
where $U \equiv (u, c, t)^T$, and $ D \equiv (d,s,b)^T$.
The usual Cabibbo-Kobayashi-Maskawa matrix is given by
\begin{equation}
V_{CKM} = V_L^{U+} V_L^D.
\label{vckm}
\end{equation}

Using unitarity of the $V^D$ and $V^U$ matrices, we obtain
flavour-changing neutral interactions~\cite{lv}
\begin{equation}
{\cal L}^{NC}_{ds}=\frac{g c_W}{2 \sqrt{3-4 s_W^2}}
\left[V^{D*}_{Lid} 
V^D_{Lis}\right] \bar{d}_L \gamma^\mu s_L 
Z'_\mu,
\label{fcnc}
\end{equation}
where $i$ denotes the number of ``different" quark family
i.e. the $ \mbox{SU(3)}_L $ quark triplet.  
 It was shown in Ref.~\cite{lv} that $i$ must be equal to 3, i.e.
the third family of quarks must be different 
from the first two.

\hs We consider the decay
\be
 K^+(p1) \rightarrow \pi^+(p2)\  \nu(k1)\  \bar{\nu}(k2),
\ee
where the symbols in parentheses stand for 
the momenta of the particles.
The one-loop effective SM  Lagrangian  for 
this process was calculated by 
Inami and Lim and other authors~\cite{il}. 
Due to family nonuniversality
 in the 3 - 3 - 1 models, the decay 
can be mediated by the $Z'$ 
at the tree level. The Feynman diagram contributing 
to the considered decay is depicted in Fig. 1

\vspace*{0.9cm}

\begin{center}
\begin{picture}(260,50)(-5,0)
\ArrowLine(20,50)(31,10)
\ArrowLine(31,10)(20,-30)
\Photon(31,10)(110,10){2}{6}
\ArrowLine(110,10)(130,50)
\ArrowLine(130,-30)(110,10)
\ArrowLine(10,50)(10,-30)
\Text(75,18)[]{$ Z' $}
\Text(14,58)[]{$K^+$}
\Text(14,-36)[]{$\pi^+$}
\Text(142,55)[]{$\nu$}
\Text(142,-35)[]{$\bar{\nu}$}
\Text(4,30)[]{$u$}
\Text(4,-20)[]{$u$}
\Text(29,-20)[]{$\bar{d}$}
\Text(30,30)[]{$\bar{s}$}
\Text(90,-80)[]{ Figure 1: Feynman diagram for 
$K^+ \rightarrow  \pi^+ \nu \bar{\nu}  $}
\Text(90,-90)[]{ in the  3 3 1 models }
\end{picture}
\end{center}
\vspace*{4cm}

The   decay amplitude is given by
\be
{\cal M}(K^+\rightarrow \pi^+ \nu  \bar{\nu})
= \frac{G_F}{ \sqrt{2}}\frac{m_W^2}{M_{Z'}^2} V^{D*}_{Lbd}
 V^{D}_{Lbs}
\langle \pi^+(p2)|\bar{s}_L \gamma_\mu d_L|K^+(p1)\rangle
\bar{\nu}_L(k1)\gamma^\mu \nu_L(k2),
\label{mt331}
\ee
where $m_W$ and $M_{Z'}$ stand for masses of the $W$ and $Z'$ 
bosons, respectively.

\hs For our initial purpose we present 
the well measured semileptonic decay
$ K^+(p1) \rightarrow \pi^0(p2)  e^+(k1)  \nu(k2)$. 
The tree-level amplitude for this process 
can be written as
\be
{\cal M}(K^+ \rightarrow \pi^0  e^+ \ \nu)
 = \frac{G_F}{ \sqrt{2}} V^{*}_{us}
\langle \pi^0(p2)|\bar{s}_L \gamma_\mu u_L|K^+(p1)\rangle
\bar{\nu}_{eL}(k1)\gamma^\mu e_L(k2).
\label{mtsm}
\ee

\hs The isospin symmetry relates hadronic matrix elements
in ~(\ref{mt331}) to ~(\ref{mtsm}) to a very 
good precision~\cite{pas}
\be
\langle \pi^+(p2)|\bar{s}_L \gamma_\mu d_L|K^+(p1)\rangle
= \sqrt{2}\langle \pi^0(p2)|\bar{s}_L 
\gamma_\mu u_L|K^+(p1)\rangle.
\label{ht}
\ee
Neglecting differences in the phase space 
of  two considered decays occuring because
 $m_{\pi^+}\neq m_{\pi^0}$ and $m_e \neq 0$, we 
 sum over three neutrino flavours and obtain
\be
\frac{Br^{rhn}(K^+ \rightarrow \pi^+ \nu   \bar{\nu})}{ 
Br(K^+ \rightarrow \pi^0  e^+  \nu)} = 6 \left( 
\frac{m_W^2}{M_{Z'}^2}\right)^2
 \left(
\frac{|V_{Lbd}^{*D}V_{Lbs}^D|^2}{|V_{us}^*|^2}\right),
\label{rat}
\ee
where the symbol  $rhn$ added to the  branching ratio
indicates the case under consideration. We now apply
the simple Fritzsch~\cite{hf} scheme as 
\begin{equation}
V^D_{ij} \approx \left( \frac{m_i}{m_j} \right)^{1/2}, 
\hspace*{1cm} i < j.
\label{hfr}
\end{equation}
     
Inserting  ~(\ref{hfr}) into ~(\ref{rat}), we obtain
\be
Br^{rhn}(K^+ \rightarrow \pi^+  \nu   \bar{\nu}) =
6 \left( 
\frac{m_W^2}{M_{Z'}^2}\right)^2
 \left(\frac{m_d m_s}{m_b^2 V_{us}^2}\right)
Br(K^+ \rightarrow \pi^0\  e^+ \  \nu) . 
\label{brrhn}
\ee
In Fig. 2 we plot $Br^{rhn}$ as a function of $M_{Z'}$,
using the  data ~\cite{pdg}
\bea 
m_W& =& 80.41\ \mbox{GeV},\  |V_{us}| = 0.2196,\ m_d = 
7\ \mbox{MeV}, \nonumber\\
m_s &=& 115\ \mbox{ MeV},  m_b = 4.3 \ \mbox{ GeV},\ 
 Br(K^+ \rightarrow \pi^0  e^+  \nu) = 4.42 \times 10^{-2}.
\eea
  The horizontal lines are the 
upper  ($4.9 \times 10^{-10}$)
and the lower   ($0.3 \times 10^{-10}$)  
experimental data~\cite{adler}.
\vspace*{1cm}

\hs From the figure we see that a lower 
bound on the $Z'$ mass is in a
range from 2.3 Tev to 4.35 TeV.
This bound is approximately twice bigger than that derived
from the mass difference of the kaon mixing 
system $\Delta m_K$ ~\cite{lv}. 
We thus arrive at  the previous conclusion again:
for the $Z'$ mass to be  relatively low,
the third family of quarks must be different from  
the other two. 
\\[0.3cm]

{\it B. The decay  in  minimal 3 -- 3 -- 1 model}\\[0.1cm]
\hs This model treats the leptons as  $\mbox{SU(3)}_L$
antitriplets~\cite{fr,dng}, with the third element being the 
antilepton  (the name of this version comes from the fact 
that no new leptons are introduced)
\be
f^{a}_L = \left( \begin{array}{c}
               e^a_L\\ -\nu^a_L\\  (e^c)^a
               \end{array}  \right) \sim (1, \bar{3}, 0).
\label{l}
\ee
Of the nine gauge bosons  $W^a (a = 1, 2, ..., 8)$ and
$B$ of $SU(3)_L$ and $U(1)_N$, four
are light: the photon ($A$), $Z$ and $W^\pm$. The remaining
five correspond to new heavy gauge bosons 
$Z'$ and $ Y^\pm$ and the doubly charged
bileptons $X^{\pm \pm}$. They are expressed  in terms of
$W^a$ and $B$  as~\cite{dng}:
\bea
\sqrt{2}\ W^+_\mu &=& W^1_\mu - iW^2_\mu ,
\sqrt{2}\ Y^+_\mu = W^6_\mu - iW^7_\mu ,\nonumber\\
\sqrt{2}\ X_\mu^{++} &=& W^4_\mu - iW^5_\mu,\nonumber\\
A_\mu  &=& s_W  W_{\mu}^3 + c_W\left(\sqrt{3}\ t_W\ W^8_{\mu}
+\sqrt{1- 3\ t^2_W}\  B_{\mu}\right),\nonumber\\
Z_\mu  &=& c_W  W_{\mu}^3 - s_W\left(\sqrt{3}\ t_W\ W^8_{\mu}
+\sqrt{1- 3\ t^2_W}\  B_{\mu}\right),\\
Z'_\mu &=&-\sqrt{1- 3\ t^2_W}\ \ W^8_{\mu}
+ \sqrt{3}\ t_W\ B_{\mu}.\nonumber
\label{apstat}
\eea

As before, the  physical states are a 
mixture of $Z$ and $Z'$,
\bea
Z_1  &=&Z\cos\phi - Z'\sin\phi,\nonumber\\
Z_2  &=&Z\sin\phi + Z'\cos\phi,\nonumber
\eea
and the mixing angle $\phi$ is also  bounded to be very small.
We can  therefore assume  $\phi \approx 0$.
Applying Eq. (4.4) in~\cite{dng} we obtain the  interactions among 
the $Z'$ and neutrinos,
\be
a'_{V}(\nu) = -  a'_{A}(\nu) = \frac{1}{2\sqrt{3}} \sqrt{1-4s_W^2}.
\label{zpnu}
\ee
One necessary vertex, namely the FCNC is given in~\cite{dpp},
\begin{equation}
{\cal L}^{NC}_{ds}=\frac{g c_W}{2 \sqrt{3(1-4 s_W^2)}}
\left[V^{D*}_{Lid} 
V^D_{Lis}\right] \bar{d}_L \gamma^\mu s_L 
Z'_\mu.
\label{fcnc2}
\end{equation} 
 
Combining (\ref{zpnu}) and~(\ref{fcnc2}) we obtain 
the  decay amplitude
\be
{\cal M}^{min}(K^+\rightarrow \pi^+ \nu  \bar{\nu})
= \frac 13 \frac{G_F}{ \sqrt{2}}\frac{m_W^2}{M_{Z'}^2} V^{D*}_{Lbd}
 V^{D}_{Lbs}
\langle \pi^+(p2)|\bar{s}_L \gamma_\mu d_L|K^+(p1)\rangle
\bar{\nu}_L(k1)\gamma^\mu \nu_L(k2).
\label{mtmin}
\ee 
From Eq.~(\ref{mtmin}) it is straightforward to obtain 
\be
\frac{Br^{min}(K^+ \rightarrow \pi^+  \nu   \bar{\nu})}{ 
Br(K^+ \rightarrow \pi^0  e^+   \nu)} = \frac 23 \left( 
\frac{m_W^2}{M_{Z'}^2}\right)^2
 \left(
\frac{|V_{Lbd}^{*D}V_{Lbs}^D|^2}{|V_{us}^*|^2}\right). 
\label{ratm}
\ee

\hs As in the previous section, in Fig. 2 
we also plot $Br^{min}$
as a  function of $M_{Z'}$.
As a consequence, the lower bound on the $Z'$ mass is in a
range from 1.25 TeV to 2.45 TeV.
This bound is  bigger than the one derived
from the mass difference of the kaon mixing 
system $\Delta m_K$ (see Dumm D. G. {\it et al} in~\cite{dpp}). 
For the $Z'$ mass to be relatively low,
the third family of quarks must be different from  the other two. 
It is worth mentioning that the branching ratio is not sensitive
to the value of $\sin^2\theta_W$, while the expression of  
$\Delta m_K$  in the minimal version is very sensitive 
due to a factor  $\frac{1}{1-4s^2_W}$.\\[0.3cm]

{\large\bf III. Conclusions }\\[0.3cm]
\hs We have considered  the rare  
kaon decay $K^+\rightarrow \pi^+ \nu  \bar{\nu}$ 
in the 3 - 3 - 1 models at the tree 
level. It was shown that in the model involving right-handed
neutrinos, the decay width is by about one order bigger than 
that in the minimal version.
 As a  result, we obtained bounds on the $Z'$ 
mass in the  range from 2.3 TeV to 4.3 TeV  in the model with 
right-handed  neutrinos and  from
1.2 TeV to 2.4 TeV  in the minimal version.  
There is a point worth noting: these mass limits
are in agreement with recent analysis~\cite{el}  showing that
there are indications of the $Z'$ in electroweak precision data.
We do hope that the new experimental data from the Collaborations
 at BNL and Fermilab will bring new indications on the Extra
neutral gauge boson $Z'$ - one of the best motivated extension
of the SM.  

\hs In this work, we consider only the CP conservating kaon decay
 $K^+ \rightarrow \pi^+ \nu \bar{\nu}$. Implications for the CP 
violating K and B decays are subjects of future studies.
\\[0.3cm]

{\large\bf Acknowledgments }\\[0.3cm]
\hs One of authors (H.N.L) thanks the APCTP for 
 financial support and hospitality extended to him.
This work is supported in part by Vietnam National 
Research programme on Natural Sciences under grant KT-04.1.2.
\\[0.3cm]

\newpage

\setcounter{figure}{1}
\begin{figure*}[thb]
\centerline{\epsfxsize=12cm\epsffile{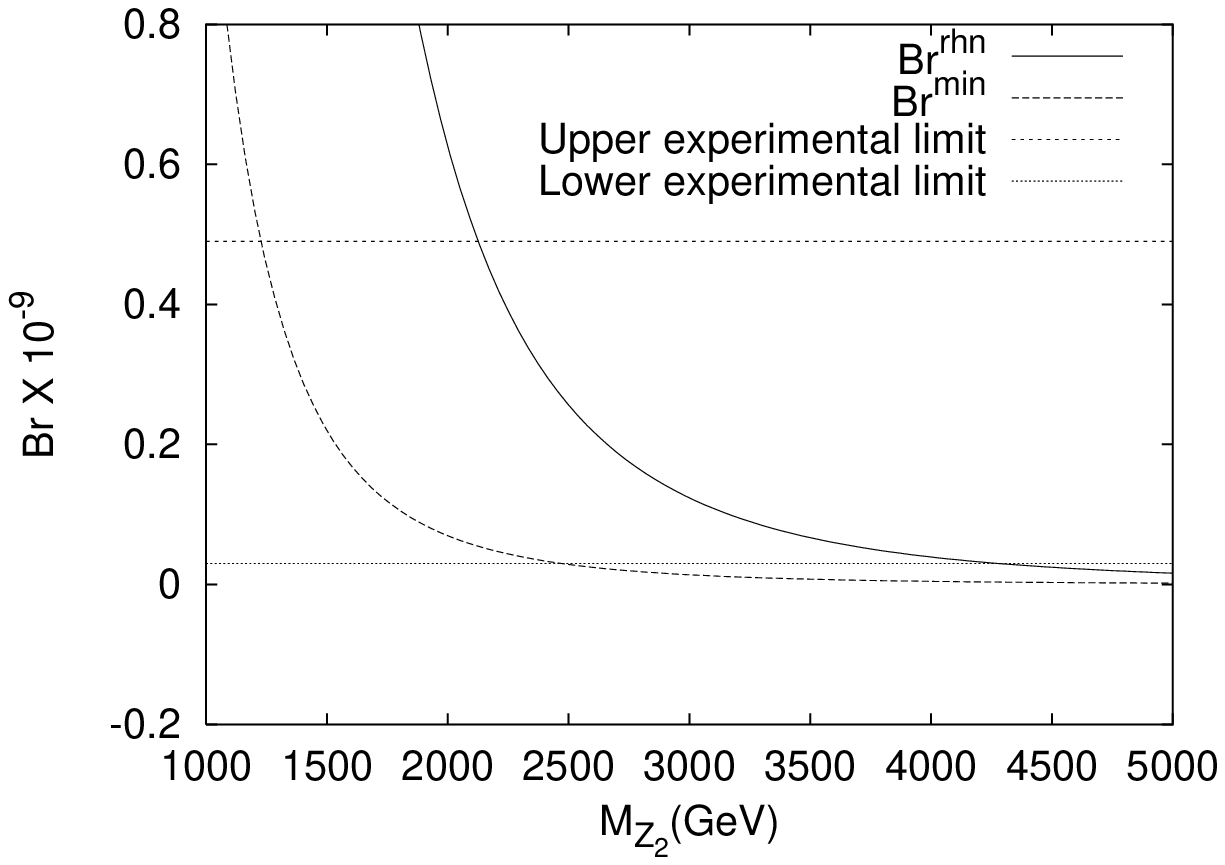}}
\caption{\label{fig2}{\em   Branching ratio (Br)  as 
a  function of $M_{Z'}$.}}
\end{figure*}
\end{document}